\documentclass[journal=nalefd,manuscript=letter,numbers]{achemso}

\DeclareUnicodeCharacter{2212}{-}
\usepackage{dcolumn}
\usepackage{bm}

\usepackage{amsmath}
\usepackage{amssymb} 
\usepackage{bm}  
\usepackage{soul}
\usepackage{mathtools}
\usepackage{upgreek}
\usepackage{array}
\usepackage{setspace}   
\usepackage{tabulary}
\usepackage{ulem}
\usepackage{xcolor}
\newcolumntype{P}[1]{>{\centering\arraybackslash}p{#1}}
\usepackage{caption}
\usepackage{booktabs}
\usepackage{makecell}
\usepackage{natbib}

\title{Confinement of Polariton Condensates in quasi-Flatband BICs in Plasmonic and Dielectric Metasurfaces}

\author{Anton Matthijs Berghuis}
\email{a.m.berghuis@tue.nl}
\affiliation{Department of Applied Physics and Science Education and Eindhoven Hendrik Casimir Institute, Eindhoven University of Technology,\\
P.O. Box 513, 5600 MB Eindhoven, The Netherlands.\\
}
\affiliation{Institute for Complex Molecular Systems-ICMS, Eindhoven University of Technology, P.O. Box 513, 5612 AJ, Eindhoven, The Netherlands}

\author{Jose Luis Pura}
\affiliation{Instituto de Estructura de la Materia (IEM-CSIC), \\
Consejo Superior de Investigaciones Científicas, \\
Serrano 121, 28006 Madrid, Spain.\\
}
\affiliation{GdS-Optronlab, Física de la Materia Condensada,\\
Universidad de Valladolid, \\
Paseo de Belén 19, 47011 Valladolid, Spain.
}

\author{Rafael P. Argante}
\affiliation{Department of Applied Physics and Science Education and Eindhoven Hendrik Casimir Institute, Eindhoven University of Technology,\\
P.O. Box 513, 5600 MB Eindhoven, The Netherlands.\\
}

\author{Shunsuke Murai}
\affiliation[kyoto]
{Department of Physics and Electronics, Graduate School of Engineering, Osaka Metropolitan University, Sakai, 5998531, Osaka, Japan
}

\author{Jos\'e A. S\'anchez-Gil}
\affiliation{Instituto de Estructura de la Materia (IEM-CSIC), \\
Consejo Superior de Investigaciones Científicas, \\
Serrano 121, 28006 Madrid, Spain.\\
}

\author{Jaime G\'omez Rivas}
\email{j.gomez.rivas@tue.nl}
\affiliation{Department of Applied Physics and Science Education and Eindhoven Hendrik Casimir Institute, Eindhoven University of Technology,\\
P.O. Box 513, 5600 MB Eindhoven, The Netherlands.\\
}
\affiliation{Institute for Complex Molecular Systems-ICMS, Eindhoven University of Technology, P.O. Box 513, 5612 AJ, Eindhoven, The Netherlands}

\begin{document}

\begin{abstract}
We investigate exciton-polariton condensation in square arrays, composed of either dielectric silicon (Si) or plasmonic silver (Ag) nanodisks, covered with a dye-doped layer. Both arrays support symmetry-protected bound states in the continuum (BICs) at normal incidence, featuring electric quadrupolar (\( Q_{xy} \)) and magnetic dipolar (\( m_z \)) characters. Due to differences in mode coupling, these BICs are split by \(\sim 10\) meV in the Si array, whereas they remain nearly degenerate in the Ag array. Simulations reveal that interference in the Ag array results in hybrid modes, \((m_z + i\tilde{Q}_{xy})\) and \((m_z - i\tilde{Q}_{xy})\), which are polarized along orthogonal directions. Interestingly, this results in similar lasing thresholds in both Si and Ag arrays, regardless of the inherent non-radiative losses of Ag, and also a confinement of the polariton condensates in the Ag array. While condensation in the Si array occurs in the \( Q_{xy} \) BIC, producing a characteristic donut-shaped far-field emission in k-space, condensation in the Ag array populates the hybrid modes, leading to a double-cross emission pattern extending over a broad range of wave vectors due to the quasi-flatband nature of this mode. As a result, the Ag array also exhibits a strong confinement along the polarization axis in real space. However, for unpolarized emission, there is a similar spatial confinement in both Si and Ag arrays. This control over the confinement of condensates could also be exploited to control interactions.  Our results highlight a novel mechanism for condensate confinement with potential applications in quantum computing and polaritonic circuitry.

\end{abstract}
\maketitle
\setstretch{1.65}


\newpage

\section{Introduction}
Exciton-polaritons are hybrid light-matter quasiparticles that can undergo stimulated relaxation into a single quantum state, resulting in coherent laser-like emission. This phenomenon is known as polariton lasing or non-equilibrium polariton condensation\cite{Kasprzak2006Bose-EinsteinPolaritons,Kena-Cohen2010Room-temperatureMicrocavity,Byrnes2014Exciton-polaritonCondensates,Ramezani2017Plasmon-exciton-polaritonLasing}. Due to their matter component, polaritons can interact with each other \cite{Kasprzak2006Bose-EinsteinPolaritons,Plumhof2014Room-temperaturePolymer,DeGiorgi2018InteractionCondensate}, as well as with other condensates \cite{Askitopoulos2015RobustCondensates}, enabling a range of nonlinear and collective phenomena \cite{Amo2009CollectiveMicrocavity,Grudinina2023CollectiveCondensate}. This property makes polariton condensates an interesting platform to explore nonequilibrium many-body physics, but also promising candidates for quantum simulation and computation \cite{Angelakis2017,Kavokin2022PolaritonComputing,Boulier2020MicrocavitySimulation}. A key requirement for such applications is the ability to control and confine the condensate in both real and reciprocal space.\\
\\
In recent years, there have been significant advances in the research of lasing from BICs and quasi-BICs,~\cite{Kodigala2017LasingContinuum,Hakala2017LasingLattice,Yang2021Low-ThresholdMetasurfaces,HuangUltrafastMicrolasers,Wu2020Room-TemperatureContinuum,Mohamed2022ControllingContinuum,Zhai2022MultimodeContinuum,Zhou2024LasingContinuum,Song2025High-QualityPolarization, Gao2024All-OpticalLattices} as well as polariton condensation~\cite{Ardizzone2022PolaritonContinuum,Riminucci2022NanostructuredContinuum,Grudinina2023CollectiveCondensate,Wu2024ExcitonTemperature,Yan2025TopologicallyMetasurfaces}. BICs offer long radiative lifetimes as a result of their symmetry-protected decoupling from free-space modes, making them ideal cavities for condensation. Furthermore, BICs have topological charges~\cite{Zhen2014TopologicalContinuum,Zhang2018ObservationSpace} and support complex polarization textures in both the near- and far-field \cite{Asamoah2022FiniteContinuum,Berghuis2024CondensationTransport}. However, the dispersion of the modes that define quasi-BICs that lead to BICs at the $\Gamma-$point (normal to the surface), and their concomitant high group velocities in extended metasurfaces often hinder the confinement of polaritons and increase the condensation threshold.~\cite{Berghuis2024CondensationTransport}
Various approaches have been developed to trap polariton condensates, including patterned microcavity arrays \cite{Jacqmin2014DirectPolaritons,Baboux2016BosonicBand} among which Lieb lattices\cite{Klembt2017PolaritonLattice,Scafirimuto2021TunableTemperature,Harder2020Exciton-polaritonsLattice}. In addition, optically defined traps \cite{Wertz2010SpontaneousCondensates,Wei2022OpticallySemiconductor} and finite-sized cavities \cite{Zhang2020TrappedMicrocavity} have been used. Another emerging method is the use of flat or quasi-flat bands in metasurfaces, which can reduce group velocities and localize modes without requiring additional patterning \cite{Nasidi2023FlatLattice,Eyvazi2025Flat-BandMetasurfaces,Do2025}. In addition, engineering the cavity geometry allows for control over the polarization state and the multipolar character of the condensate, enabling further tuning of its interaction properties.\\
\\
In this manuscript, we demonstrate polariton condensation into a bound state in the continuum (BIC) in a quasi-flat band supported by a plasmonic silver (Ag) metasurface. By combining real-space and momentum-space imaging, we show that the condensate exhibits strong confinement in real space along one axis for a single polarization state of the condensate. Although a dielectric silicon (Si) metasurface with similar geometry also supports a BIC, this mode is more dispersive, resulting in a condensate that is delocalized in real space but exhibits a narrow distribution in momentum space. We further provide a physical understanding of the distinct dispersions observed in the Ag and Si arrays. By performing a multipolar decomposition of the BICs, we show that the difference arises from the character of the modes in which the condensates are formed. In the Si array, condensation takes place in a quadrupolar mode, while in the Ag array, condensation occurs in two orthogonal modes $m_z \pm i\tilde{Q}_{xy}$.
This different hybridization determines the polariton dispersion and polarization properties. Our results provide new insight into the design of metasurfaces for polariton confinement and set the stage for future studies of polariton–polariton interactions in metasurfaces.

\section{Results}
\begin{figure}[H] 
  \begin{center}
    \includegraphics[width=.95\textwidth]{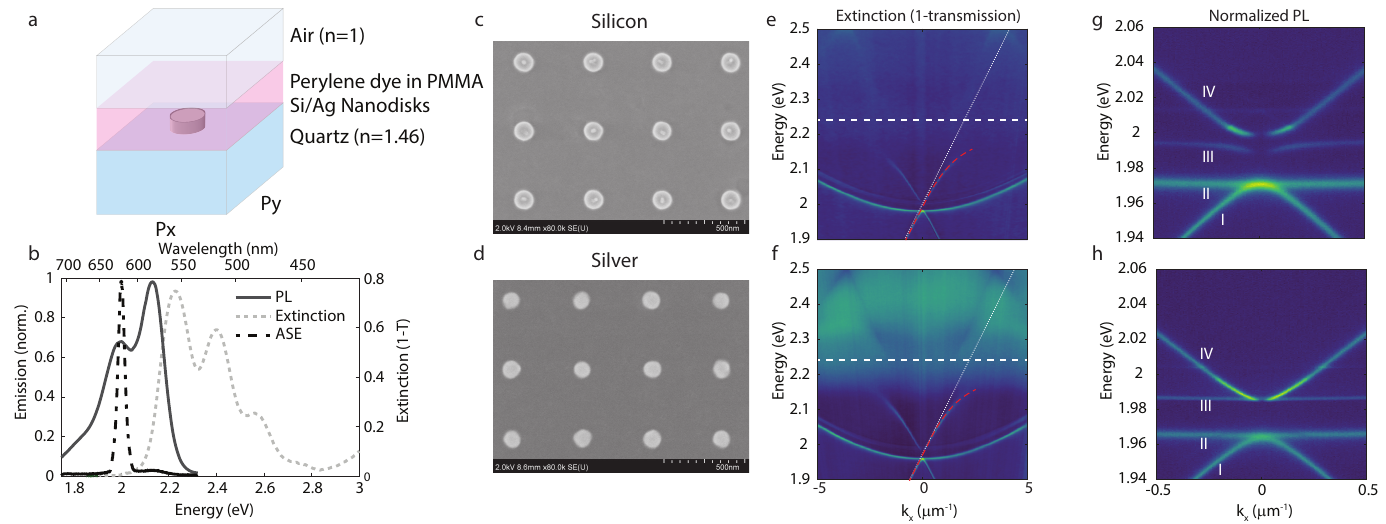}
    \end{center}
  \caption{\setstretch{1.65} a) Schematic geometry of the samples. Nanodisks from either Ag or Si are fabricated on a quartz substrate using e-beam lithography in a square array. The nanodisks are covered with a layer of PMMA with a thickness of 220 nm doped with perylene dye molecules at a concentration of 30 wt\%. (b) Extinction (light-gray-dashed curve), emission (solid-black curve), and amplified spontaneous emission (dark-gray-dash-dotted curve) of the bare molecules. (c) SEM image of the Si and (d) Ag nanoparticle arrays. (e) and (f) Dispersion obtained by extinction measurements of the Si and Ag metasurfaces, respectively. Both systems are in the strong coupling regime, evidenced by the anti-crossing of the metasurface mode with the exciton resonance energy at 2.24 eV (indicated with the horizontal white-dashed line). (g) and (h) Close-ups of the photoluminescence (PL) from the strongly coupled metasurfaces near k$_x$=0. Both, the Si and Ag metasurfaces show a BIC at normal incidence.}
  \label{fgr:Figure1}
\end{figure}

We have designed two square arrays, both with a period 420 nm, consisting of disks of either Si or Ag nanodisks on top of a quartz substrate using electron beam lithography (see Supporting Information (SI) Section S1). A single unit cell is schematically depicted in Fig.~\ref{fgr:Figure1} a. To compensate for the intrinsic larger scattering cross section of metal (plasmonic) nanoparticles compared to dielectric (Mie) particles, the Ag nanodisks (diameter = 50 nm, height = 60 nm) have smaller dimensions than their Si counterpart (diameter = 90 nm, height = 90 nm), as also visible in the SEM images of the Si and Ag arrays in Figs.~\ref{fgr:Figure1} c and d, respectively. These dimensions were chosen such that the quality factor ($Q$) of the bright surface lattice resonance of both structures yields approximately the same value (Si: Q$\sim$200, Ag: Q$\sim$150, based on simulations (see SI Sections S2). Additionally, the Ag nanoparticle arrays have a 2 nm $TiO_2$ adhesion layer and a 20 nm $Al_2O_3$ protective layer to prevent oxidation (see SI section for fabrication details).\\
\\
As in our previous work\cite{Berghuis2023RoomContinuum}, the nanoparticle arrays are covered with a $\sim$ 220 nm thick layer of dye-doped PMMA (30\% of perylene dye [N, N'-Bis(2,6-diisopropylphenyl)-1,7- and -1,6-bis (2,6-diisopropylphenoxy)-perylene-3,4:9,10-tetracarboximide] in PMMA). We use an organic dye because the low dielectric screening results in tightly bound Frenkel excitons that are stable at room temperature. This perylene dye has a large quantum yield and high solubility in PMMA without forming aggregates\cite{Ramezani2017Plasmon-exciton-polaritonLasing}. The absorption and emission spectra are shown in Fig. \ref{fgr:Figure1} b. where the light gray dashed curve is the absorption spectrum, the solid black curve is the emission spectrum and the dark gray dash-dotted curve the amplified spontaneous emission, under intense pulses laser excitation (data reproduced from Ref.[\citenum{Berghuis2023RoomContinuum}]). 

The extinction of the samples is characterized in a Fourier microscope (see SI section S3 for setup details), allowing to measure the band structure up to $37^\circ$ with respect to normal incidence. The Fourier plane is imaged onto a slit and dispersed in energy, resulting in the band structures along the in-plane wave vector k$_x$, as shown in Fig.~\ref{fgr:Figure1} e for the Si array and Fig.~\ref{fgr:Figure1} f for the Ag array. At normal incidence, the surface lattice resonances (SLRs) at $\sim$ 2 eV are red detuned from the Perylene main exciton resonance at 2.24 eV (indicated with the white dashed lines in Figs. \ref{fgr:Figure1} e and f). However, at larger wave vectors, where the energy of the SLR approaches the exciton energy, a clear coupling is visible by the avoided crossing as the SLR bends away from its linear dispersion, indicated with the diagonal white-dotted and red-dashed lines in Figs. \ref{fgr:Figure1} e and f. It should be noted that the unpolarized dispersions of the Si and Ag samples appear similar in Figs. \ref{fgr:Figure1} e and f, due to the tuning of the particle sizes. When we, however, take a closer look at the modes around $k_x \sim 0$, we observe an interesting difference: In Figs. \ref{fgr:Figure1} g and h, the emission from the metasurface close to normal incidence is plotted. In these plots, four different modes are visible. The modes labeled by the Roman numbers I and II are the dipolar SLRs. These modes are bright for all wave vectors and correspond to two dipolar modes along the y- and x-directions, respectively. Both modes are similar for the Si and Ag arrays. Modes III and IV are symmetry-protected BICs. Due to the vanishing of the overlap of the mode profiles with the free space radiation at $k_x=0$, their extinction goes to zero at this wave vector. For the Si arrays, we showed in a previous article \cite{Berghuis2024CondensationTransport} that BIC III corresponds to a magnetic dipolar (MD) mode in the z direction (out of plane), while BIC IV corresponds to an electric quadrupolar (EQ) mode in the xy-plane. In the Si metasurface, there is a splitting between the magnetic and quadrupolar modes of approximately 10 meV. In contrast, these modes are (almost) degenerate in the Ag metasurface resulting in hybrid magnetic-quadrupolar modes ($m_z \pm i\tilde{Q}_{xy}$ as will be explained in more detail below and in SI Section S4, where the far field polarization of the different modes is shown). \\

To shed light on the underlying physics, we carried out numerical calculations based on the free software SMUTHI (scattering by multiple particles in thin-film systems). The details on the simulations and simulated dispersions can be found in section S5 of the SI~\cite{Egel2021SMUTHI:Interfaces}. SMUTHI is based on the T-matrix method to account for the single particle scattering and on the scattering-matrix method for the propagation through a layered medium. The dipolar/quadrupolar nature of the modes is confirmed with SMUTHI by calculating extinction bands up to the dipolar ($l=1$), quadrupolar ($l=2$), and octupolar ($l=3$) order, the latter only to discard higher-order contributions. The condensation mode in the case of the Si metasurface is attributed to the EQ BIC, spectrally separated from all the MD modes. This separation in the case of Si appears because the fields of the magnetic mode have a larger contribution inside the particle compared to the Ag array, while the fields of the quadrupolar lie mostly outside the particle for both Si and Ag metasurfaces (SI Section S6). In the case of the Ag metasurface, the calculations confirm the spectral overlap of the EQ and MD modes. Remarkably, SMUTHI allows for the identification of two related BICs separated by $\sim 10^{-3}$ eV, due to the fact that the multipoles of cylinders are very accurately accounted for. In practice, both modes ought to be mixed due to the obvious imperfections of the fabricated nanodisks. This will be further discussed below in connection with the condensate vortex emission shape and polarization.\\

\begin{figure}[H] 
  \begin{center}
    \includegraphics[width=.5\textwidth]{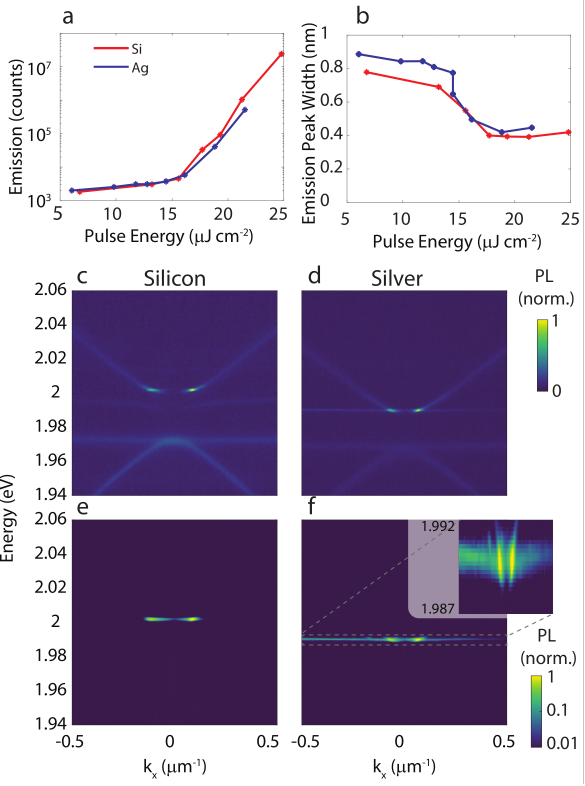}
    \end{center}
  \caption{\setstretch{1.65} (a) Condensation threshold curves for the Si (red curve) and Ag (blue curve) metasurfaces. (b) Linewidths of the emission from the condensates at k$_x=0.05 \upmu $m$^{-1}$ as a function of pump fluence for Si (blue curve) and Ag (red curve). (c)-(d) Emission dispersion at threshold (P=P$_{th}$) for the Si and Ag metasurfaces, respectively. (e)-(f) Emission dispersion at P$\sim$1.3P$_{th}$ on a logarithmic color scale. The dispersions are plotted as a function of k$_x$ with k$_y$=0.
  }
  \label{fgr:Figure3}
\end{figure}

When excitons are generated with sufficient laser power ($\lambda=400$ nm, 1 kHz, 100 fs pulse from Astrella, Coherent), they can relax into a single mode, forming a non-equilibrium polariton condensate or polariton laser. The increase of the emission from the BIC upon reaching the condensation threshold is very similar for the Ag and Si arrays, as shown in Fig. \ref{fgr:Figure3} a, where the emission intensity at $k_x=0.05$ $\upmu$m$^{-1}$ is plotted as a function of the absorbed fluence. Both systems also show a similar narrowing of the emission linewidth upon reaching threshold at a pump fluence around 14 $\upmu$J cm$^{-2}$, resulting in a linewidth of $\sim$ 0.4 nm for a  excitation spot size with a diameter of 120 $\upmu$m.\\
\\

In the Si metasurface, the excitons relax into the quadrupolar mode, as observed in the dispersion at the condensation threshold (P$_{th}$) in Fig. \ref{fgr:Figure3} c. In this figure, the polaritons that have started to condense into the BIC are leaking from the condensate at two discrete points in reciprocal space. The condensation takes place into the quadrupolar BIC, due to its longer lifetime compared to the lifetime of the magnetic BIC, which enables a faster buildup of the polariton condensate before the polaritons decay (Figure S6g and Ref. [\citenum{Berghuis2023RoomContinuum}]). The condensate emission at the threshold in the Ag metasurface appears very similar, as shown in the emission at two discrete wave vectors in Fig. \ref{fgr:Figure3} d. The main differences are that the emission is at slightly lower energy and smaller wave vectors compared to the Si array, due to the different coupling between the lattice mode and the localized resonances of the particles.\\

\begin{figure}[H] 
  \begin{center}
    \includegraphics[width=.8\textwidth]{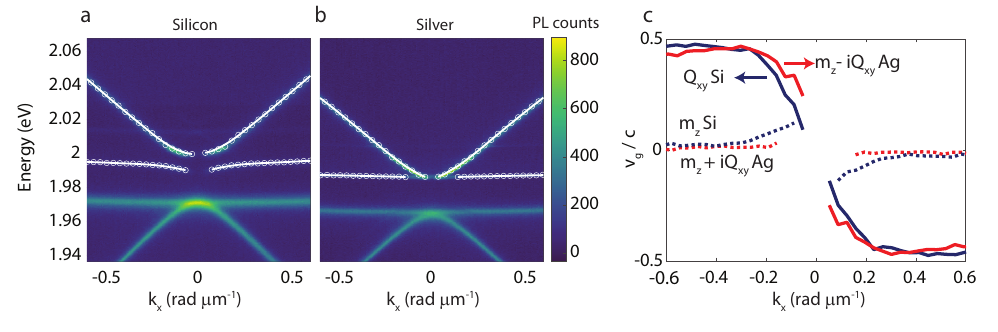}
    \end{center}
  \caption{\setstretch{1.65}The dispersion of the Si array (a) and Ag array (b) with the peak maxima of the (quasi) BICs indicated with the white circles. The white curves are a smoothed curve through the data used to calculate the derivative. (c) Group velocity of the magnetic dipolar and electric quadrupolar modes in the Si array represented by the blue solid and dashed curves, respectively, and the group velocity of the hybrid modes ($(m_z + i\tilde{Q}_{xy})$ and  $(m_z - i\tilde{Q}_{xy})$) in the Ag array by the red solid and dashed curves, respectively.
  }
  \label{fgr:GroupVelocities}
\end{figure}

Due to the near degeneracy of the two BICs, a pronounced difference between the Si and Ag metasurfaces becomes apparent when the emission above threshold (P=1.3 P$_{th}$) is compared. To emphasize this difference, we plot the emission on a logarithmic scale for the Si and Ag metasurfaces in Figs.~\ref{fgr:Figure3} e and f, respectively. In addition to the emission at the two discrete wave vectors, the condensate in the Ag metasurface also emits from a quasi-flat band that follows the dispersion of the magnetic BIC, as shown in the inset of Fig.~\ref{fgr:Figure3} f. While this mode is also a BIC, a non-zero emission is visible at k$_x$=0, which is explained by the width of the spectrometer slit, collecting emission from wavevectors slightly larger than k$_y$=0, where the mode is not dark. 
The intensity of the emission from this flatband at k$_y$=0 is relatively low. However, at slightly larger values of k$_y$ its emission dominates, as discussed below in Fig. \ref{fgr:Figure4} and in SI section S7.
The appearance of this flat band implies a low group velocity of the polaritons, which is confirmed when we calculate the group velocities of the different modes from their dispersion. We fit the peaks of the dispersion of the magnetic and quadrupolar modes in the Si array in Fig. \ref{fgr:GroupVelocities} a, and those of  the hybrid $(m_z + i\tilde{Q}_{xy})$ and  $(m_z - i\tilde{Q}_{xy})$ modes in the Ag array in Fig. \ref{fgr:GroupVelocities} b, where the maxima are indicated with the white circles. From the resulting dispersion, we obtain the group velocities of the modes as $v_g=\frac{d\omega}{dk}=\hbar\frac{dE}{dk}$. Since noise in the peak position is amplified when taking the derivative, the peak positions are smoothed by a moving average, as shown with the white curves in Figs. \ref{fgr:GroupVelocities} a and b. The group velocity of the hybrid $(m_z + i\tilde{Q}_{xy})$ mode in the Ag array (red dashed curve in Fig. \ref{fgr:GroupVelocities} c) is clearly lower (along k$_x$) than the group velocity of the quadrupolar (Q$_{xy}$ mode) in Si (blue solid curve in Fig. \ref{fgr:GroupVelocities} c). At k$_x$=0.2 rad $\upmu$m$^{-1}$, the group velocities for the Ag array are 0.45 c, with c the speed of light in vacuum, for ($m_z - i\tilde{Q}_{xy}$) and 0.012 c for ($m_z + i\tilde{Q}_{xy}$), whereas for the Si array, the group velocities are 0.39 c for the Q$_{xy}$ mode and 0.08 for the m$_z$ mode (blue dashed curve in Fig. \ref{fgr:GroupVelocities} c). So, at this wave vector, the group velocity of the quasi-flat band in the Ag array is almost 40 times slower than that of the quadrupolar mode in the Si array, implying a much stronger localization of the polariton condensate in quasi-flatband of the Ag metasurface.\\

To further study the condensates, we have measured the emission in Fourier and real space. Since the emission profiles in real and Fourier space are related through the Fourier transform, the extent of the emission $\Delta k$ (the full width at half maximum of the peaks) in Fourier space is inversely proportional to the propagation length of the polaritons $L_p\propto \frac{1}{\Delta k}$ (derivation in SI Section S8). Therefore the localization should be higher for the modes with a larger spread in Fourier space. To verify this, we excite the metasurfaces with a Gaussian beam, which defines the shape of the emission at low excitation fluences. Below threshold, the fluorescence is dominated by incoherent polaritons, as shown in Figs.~\ref{fgr:Figure4} a and b for the Si and Ag metasurfaces, respectively. When the threshold is reached, the emission profile becomes dominated by the coherent emission from the condensate, leading to very different properties. For the Si metasurface, the fluorescence (real space) originates from a donut shape area with four lobes (Fig. \ref{fgr:Figure4} c), and the far-field emission measured in the Fourier plane resembles a vortex with no emission in the normal direction due to the quadrupolar BIC (Fig. \ref{fgr:Figure4} d) \cite{Asamoah2022FiniteContinuum}. The emission from the condensate in the Ag metasurface is very different. The far-field emission shows a double cross that spreads over much larger wave vectors (Fig.~\ref{fgr:Figure4} e). The flat band that was observed in the dispersion in Fig. \ref{fgr:Figure3} f, agrees with these extended bands.
Due to the distribution of the condensate over a wide range of wave vectors and the slow group velocity, we would expect a confined condensate in real space. This confinement is, however, not directly clear from the real space image of the condensate, resembling an extended cross shape as shown in Fig.~\ref{fgr:Figure4} f. \\

To explain this last observation, we have added a polarizer in the detection path to distinguish between the different polarizations of the condensate. When the polarizer is oriented along the y-direction, the emission from the Si metasurface only originates from two lobes along the x-direction (Fig.~\ref{fgr:Figure4} g). For this polarization, there is no emission at $k_y=0$ (Fig.~\ref{fgr:Figure3} h), in agreement with the polarization of an electric quadrupole vortex. For the Ag metasurface, the emission originates only from the bands parallel to the $k_y$ axis (Fig.~\ref{fgr:Figure4} i). The y-polarized emission has very confined momentum along $k_x$, resulting in a low spatial confinement along this direction (Fig.~\ref{fgr:Figure4} j). However, the momentum of the condensate takes a broad range of values along $k_y$, resulting in the spatial confinement of the condensate along the y-direction. This explains why, without a polarizer, the condensate does not appear spatially confined. A further analysis of the group velocities of all modes is shown in the SI Section S9.\\

To quantify the confinement of the condensates, we take the profile of the excitation spot and the condensates in real space, along the white dashed lines in Figs.~\ref{fgr:Figure4} a,b,g, and j. The cross sections are shown in Fig.~\ref{fgr:Figure4} k along the x-axis below threshold (black curve), and above threshold (red curve) for y-polarized emission. Upon condensation, the shape of the emission changes into two peaks, with a total width slightly larger than below threshold (the FWHM is $\sim 215$ $\upmu$m below threshold and $\sim 290$ $\upmu$m above threshold for the two peaks together). However, the individual peaks have a width of approximately one third of this value, i.e. $\sim 100$ $\upmu$m. Along the y-direction, the spot is actually a bit smaller above threshold than below, reducing from $\sim 215$ $\upmu$m to $\sim 160$ $\upmu$m, as shown in  Fig. \ref{fgr:Figure4} l. This reduction can be understood by the non-linear dependence of the condensate population with the exciton density. It is therefore much more efficient to form the condensate in the center of the excitation region, where this density is larger than at the edges. Above the threshold, the emission is dominated by the condensate, and the overall width reduces.\\ 

For the Ag metasurface, the expansion of the condensate above threshold along the x-direction is even larger than for the Si array, increasing the FWHM from $\sim 160$ $\upmu$m below threshold to $\sim 250$ $\upmu$m above (see Fig.~\ref{fgr:Figure4} m). This expansion is not surprising since the Fourier image shows the emission at a well-defined wave vector along x (Fig.~\ref{fgr:Figure4} i), implying larger uncertainty in the position. The emission profiles for the Ag array along the y-direction are more interesting. We see that above the threshold the size sharply decreases from $\sim 160$ $\upmu$m to $\sim 50$ $\upmu$m (see Fig.~\ref{fgr:Figure4}n). The non-linear process of condensation, combined with the excellent spatial confinement, explains this strong contraction of the emission profile. The excellent confinement is explained by the low group velocity associated with the flat band and broad distribution of the condensate's wave vectors along k$_y$. An overview of the condensate and excitation profiles is given in Table \ref{tab:condensate_sizes}. These results agree with the group velocities of the condensates in the x- and y-directions for the Ag and Si metasurfaces, as discussed in Fig. \ref{fgr:GroupVelocities}. \\

\newpage

\begin{figure}[H] 
  \begin{center}
    \includegraphics[width=.9\textwidth]{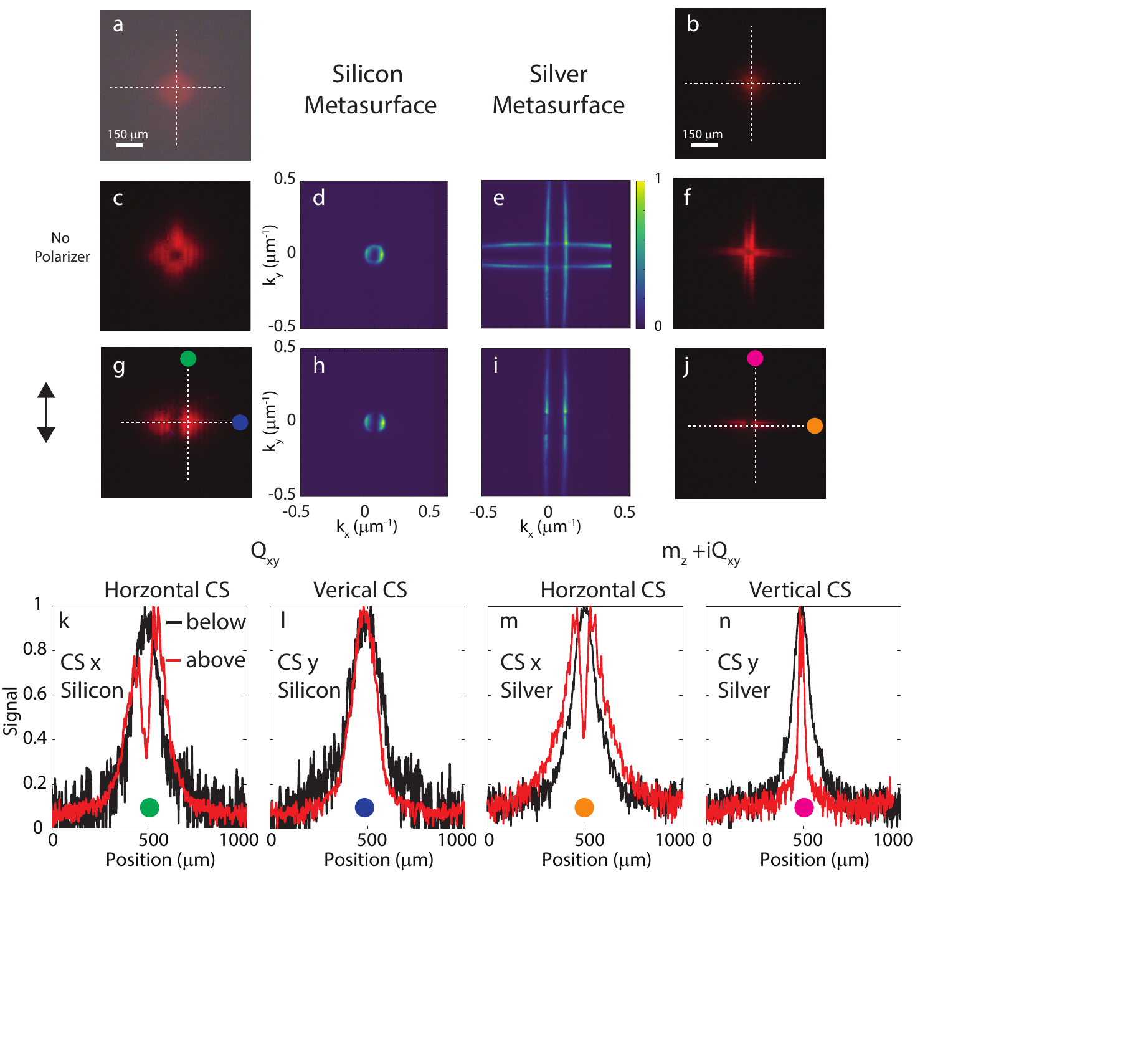}
    \end{center}
\caption{\setstretch{1.6} To fully investigate the condensation mode of the Ag and Si arrays, we map the condensate in both real space (red plots) and reciprocal space (green/blue plots) using different orientations of the polarizer in the detection. (a) and (b) show the area of emission below threshold. (c-f) show the emission from the condensate without a polarizer. (g-j) show the vertical polarized fraction of the emission. The emission for the other polarizations are plotted in the SI, Figure S11. The cross sections of the real-space dimensions of the condensate (along the white dashed lines) are plotted in (k-n) and the resulting widths corresponding to the confinement of the condensates are given in Table \ref{tab:condensate_sizes}}
  \label{fgr:Figure4}
\end{figure}

\begin{table}[h]
    \centering
    \begin{tabular}{lccc}
        \toprule
        & \makecell{\textbf{Excitation} \\ \textbf{Diameter}} 
        & \makecell{\textbf{Condensate size} \\ \textbf{for Vertical Polarization}}
         & \makecell{\textbf{ratio} \\ \textbf{(Condensate/Excitation)}} \\
        \midrule
        \multicolumn{4}{l}{\textbf{Si}} \\
        x & 215 $\upmu$m & 290 $\upmu$m & 1.35 \\
        y & 215 $\upmu$m & 160 $\upmu$m & 0.74 \\
        \midrule
        \multicolumn{4}{l}{\textbf{Ag}} \\
        x & 160 $\upmu$m & 250 $\upmu$m& 1.56 \\
        y & 160 $\upmu$m& 50 $\upmu$m & 0.31 \\
        \bottomrule
    \end{tabular}
    \caption{Excitation diameter, condensate size for vertical polarization, and ratio for Si and Ag cavities.}
    \label{tab:condensate_sizes}
\end{table}

\begin{figure}[H] 
  \begin{center}
   \includegraphics[width=.65\textwidth]{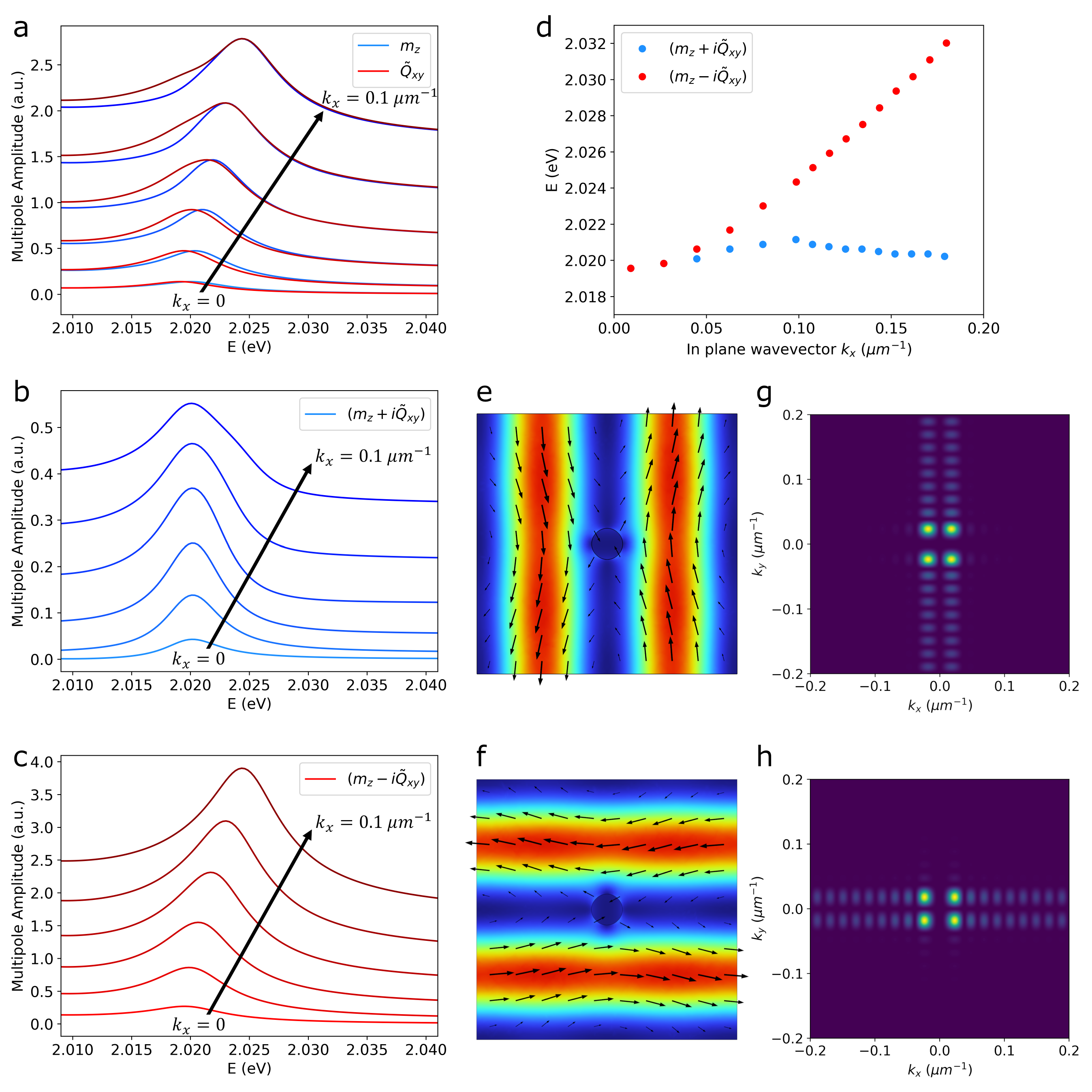}
    \end{center}
  \caption{\setstretch{1.65} (a) Multipolar decomposition of the BIC in the Ag array, showing two contributions of a magnetic dipole along the z-axis, $m_z$ (blue curves), and an electric quadrupole in the $xy-$plane, $\tilde{Q}_{xy}$ (red curves). The modes are mixed, and the decomposition is not appropriate. (b) and (c) Multipolar decomposition of the BIC in the Ag array using the \{$(m_z + i\tilde{Q}_{xy})$, $(m_z - i\tilde{Q}_{xy})$\} basis. The two modes are now clearly separated, with the $(m_z + i\tilde{Q}_{xy})$ mode accounting for the flatband. (d) Energy dispersion of the $(m_z + i\tilde{Q}_{xy})$ (blue dots) and $(m_z - i\tilde{Q}_{xy})$ modes (red dots) as a function of the in-plane wavevector $k_x$. Near field distributions of the $(m_z + i\tilde{Q}_{xy})$ (e) and $(m_z - i\tilde{Q}_{xy})$ (f) modes. Theoretical far-field emission of the $(m_z + i\tilde{Q}_{xy})$ (g) and  $(m_z - i\tilde{Q}_{xy})$ (h) modes for an array with N = 750 unit cells. Note that the y-polarized emission from panel (g) reproduces the experimental emission from Fig.~\ref{fgr:Figure4}i.
  }
  \label{fgr:Figure5}
\end{figure}

We have further analyzed the condensation mode using COMSOL Multiphysics, calculating the multipolar decomposition of the BIC mode near the $\Gamma$ point. Figure~\ref{fgr:Figure5}a shows the contributions of two non-degenerate modes for the Ag array with different multipolar behaviour: a dipolar magnetic contribution along the z-axis, $m_z$, and an electric quadrupole in the $xy-$plane, $Q_{xy}$. These two modes also appear in the Si arrays, but in that case, they are clearly separated in frequency. To directly compare $Q_{xy}$ with $m_z$, we defined the magnitude $\tilde{Q}_{xy} = \frac{\omega}{6}Q_{xy}$, which has units of magnetic moment. As shown in Figure~\ref{fgr:Figure5}a, both modes overlap and are dispersive, and the decomposition does not reproduce the experimental results. Therefore, we need to find base that reproduces the different dispersion of the two modes.
\\

The radiation fields, $\vec{E}_m$ and $\vec{E}_Q$, of a magnetic dipole $\vec{m}$ and an electric quadrupole $\overleftrightarrow Q$ are, respectively \cite{jackson_classical_1999},
  \begin{equation}
      \vec{E}_m = -\frac{Z_0 k^2}{4\pi}\frac{e^{ikr}}{r}(\hat{r}\times\vec{m}), \quad \vec{E}_Q = \frac{ik^3}{24\pi\varepsilon_0}\frac{e^{ikr}}{r}\left[\left(\hat{r}\times\vec{Q}(\hat{r})\right)\times\hat{r}\right] \;,
  \end{equation}
where $\hat{r}$ is the unit vector of the direction of observation, $Z_0 = \sqrt{\mu_0 c}$ is the impedance of free space, $\vec{Q}(\hat{r}) = \overleftrightarrow Q \cdot \hat{r}$ is the projection of the electric quadrupole tensor along $\hat{r}$, and the harmonic time dependence is implicit. If we consider the components of the relevant multipoles, $m_z$ and $Q_{xy}$ in this case, these relations can be further simplified to
  \begin{equation}
      \vec{E}_{m_z} = -\frac{Z_0 k^2}{4\pi}\frac{e^{ikr}}{r}m_z \begin{pmatrix}r_y\\-r_x\\0\end{pmatrix}, \quad \vec{E}_{Q_{xy}} = \frac{ik^3}{24\pi\varepsilon_0}\frac{e^{ikr}}{r}Q_{xy}\begin{pmatrix}r_y(1-2r_x^2)\\r_x(1-2r_y^2)\\-r_xr_yr_z\end{pmatrix} \approx \frac{ik^3}{24\pi\varepsilon_0}\frac{e^{ikr}}{r}Q_{xy}\begin{pmatrix}r_y\\r_x\\0\end{pmatrix} \;,
  \end{equation}
where $r_i$ are the cartesian components of $\hat{r}$, and the last approximation is valid for small angles, i.e. close to the $\Gamma$ point, which is the region of interest. These two equations result in the far-field vortices of $m_z$ and $Q_{xy}$. By looking at the prefactors of both equations and using $\tilde{Q}_{xy} = \frac{\omega}{6}Q_{xy}$ the total radiation field can be written as
    \begin{equation}\label{eq:far-field}
      \vec{E} = \vec{E}_{m_z} + \vec{E}_{Q_{xy}} = -\frac{Z_0 k^2}{4\pi}\frac{e^{ikr}}{r}\begin{pmatrix}(m_z - i\tilde{Q}_{xy})r_y\\ -(m_z + i\tilde{Q}_{xy})r_x\\0\end{pmatrix} \;.
  \end{equation}

According to this derivation, instead of using the modes \{$m_z$, $\tilde{Q}_{xy}$\} as a base, we can use a linear combination of these states: \{$(m_z + i\tilde{Q}_{xy})$, $(m_z - i\tilde{Q}_{xy})$\}. The results of the multipolar decomposition in the new basis are shown in Figures \ref{fgr:Figure5}b and c. Within this picture, the two orthogonal modes become clearly independent, with $(m_z + i\tilde{Q}_{xy})$ (Figure \ref{fgr:Figure5}b) corresponding to the non-dispersive flatband, and $(m_z - i\tilde{Q}_{xy})$ (Figure \ref{fgr:Figure5}c and SI Sec S10) the mode with a parabolic dispersion. The energy dispersion of both modes as a function of the in-plane wavevector $k_x$ is presented in Figure \ref{fgr:Figure5}c. Figures \ref{fgr:Figure5}e and f show the near-field distribution of both modes at the $\Gamma$ point on a plane crossing the Ag nanodisk at its center. Finally, Figures \ref{fgr:Figure5}g and h show the theoretical far-field radiation computed from Eq. \ref{eq:far-field}, which completely resemble the experimental measurements of Figure \ref{fgr:Figure4}i and together they reproduce the double cross pattern.


\section{Conclusions}
We have designed two similar square nanoparticle arrays covered with a dye-doped layer, made of either dielectric Si nanodisks or plasmonic Ag nanodisks. Both arrays supported an electric quadrupolar ($Q_{xy}$) and a magnetic dipolar ($m_z$)  BIC, at k=0, and show similar threshold values for lasing, regardless of the inherent losses of the plasmonic Ag nanodisks. Also, due to the different distribution of the near fields, the two BICs were split by $\sim$10 meV in the Si array, while being practically degenerate in the Ag array. Simulations showed that the interference of  $Q_{xy}$ and $m_z$ in the Ag array leads to two possible hybrid modes, $(m_z + i\tilde{Q}_{xy})$ and  $(m_z - i\tilde{Q}_{xy})$,  polarized  along either the $y$ or $x$ axis, respectively. For the Si array, we showed that condensation takes place into the $Q_{xy}$ BIC, with its expected vortex in the far-field emission. On the other hand, the Ag array condensed into the hybrid $(m_z + i\tilde{Q}_{xy})$ and $(m_z - i\tilde{Q}_{xy})$ orthogonally polarized modes. The far-field emission exhibited a double cross-shaped pattern, extending over much larger wave vectors in the far field. This lack of confinement of the wave vectors was caused by the (partially) flat band nature of the $(m_z + i\tilde{Q}_{xy})$ mode.
Conversely, the spatial confinement of the condensate in the Ag array along the polarization direction was highly effective, yielding a condensate size five times smaller along the confined axis under symmetric Gaussian excitation. This method of condensate confinement could prove valuable for investigating interactions between multiple condensates, with potential applications in quantum computing..

\begin{acknowledgement}

J.L.P. and J.A.S.G. acknowledge financial support from the grants TED2021-130786B-I00 and PID2022-137569NB-C41 (LIGHTCOMPAS), PID2021-126046OB-C22, funded by\\ MCIN/AEI/10.13039/501100011033, “ERDF A way of making Europe”, and European Union NextGenerationEU/PRTR. J.L.P. also acknowledges the financial support from a Margarita Salas contract CONVREC-2021-23 (University of Valladolid and European Union NextGenerationEU). SM acknowledges the Bilateral Joint Research Project (JPJSBP120239921) from Japan Society for the Promotion of Science.  Chat GPT 5.0 has been used for language editing of the abstract and introduction section. 

\end{acknowledgement}

\section*{Supporting Information}

SI 1: FDTD simulations of quality factors.
SI 2: Sample fabrication.
SI 3: Fourier Microscope setup.
SI 4: Far-Field polarization of the modes.
SI 5: Band structure resolved full reciprocal space - Experiment and Simulations.
SI 6: Near-fields and Q-factors of the BICs.
SI 7: Flat band lasing at k$_y>0$.
SI 8: Relation between propagation length and mode linewidth in Reciprocal space
SI 9: Group velocities in full reciprocal space.
SI 10: Calculated dispersions of different multipole components in the Ag array.
SI 11: Condensation at different energy in Ag array.
SI 12: Real space and reciprocal space images for different orientations of the detection polarizer

\section*{Author Contributions}
JGR and AMB designed the optical experiments. AMB and RPA conducted the experiments and analyzed the experimental results. JLP and JSG developed the theoretical model and performed the Comsol and SMUTHI simulations. SM fabricated the samples. AMB, JLP, JSG and JGR wrote the manuscript. All authors accepted the responsibility for the content of the manuscript and consented to its submission, reviewed all the results, and approved the final version of the manuscript. 
The authors state no conflict of interest.
\section*{Data Availability}
The datasets generated and/or analysed during the current study are available from the corresponding author upon reasonable request. 


\bibliography{references.bib}

\end{document}